# Synthesis, crystal growth and superconducting properties of Fe-Se system

A. E. Karkin, A. N. Titov, E. G. Galieva, A. A. Titov and B. N. Goshchitskii
*Institute of Metal Physics, Ural Branch of the Russian Academy of Sciences, 620219, Ekaterinburg, Russia*

Single crystals FeSe*x* were grown in an evacuated sealed quartz tube using polycrystalline material by gas-transport reaction with $I_2$ as a gas-carrier. Crystals with hexagonal-prismatic, tetragonal-prismatic, planar-square and hexagonal faceting 0.1 to 0.5 mm in size were obtained. The electronic transport and magnetic properties measurements of FeSe*x* single crystal exhibit an onset of superconducting transition $T_c$ at up to 24 K.

$FeSe_x$ is the simplest among the new Fe-based superconducting materials. It presents tetragonal iron selenide consisting of layers of edge-sharing FeSe tetrahedra, found to be superconducting at 8.5 K [1, 2], with superconductivity increasing at up to 15 K with S and Te substitutions [3]. In addition, with increase in pressure up to 8.9 GPa, the $T_c$ of $FeSe_x$ increases to 36.7 K [4, 5, 6]. This comparative chemical simplicity of FeSe might make it a perfect candidate for study of the interplay of structure and superconductivity in this superconducting family. Dispute has arisen in the literature regarding its true stoichiometry, and this suggests that many of the results reported to-date may have been obtained on multiple-phase materials. The report describing the discovery of superconductivity in tetragonal $FeSe_x$ [1] suggested that superconductivity was only observed in samples with significant selenium deficiency, and that stoichiometry of the superconducting phase was between $FeSe_{0.82}$ and $FeSe_{0.88}$. On the other hand, it was asserted in several studies that phase-pure materials can only be produced when the starting mixtures are very close to being stoichiometric [7, 8, 9]. Furthermore, new neutron powder diffraction data [10, 11] have shown that $x$ in superconducting $FeSe_x$ is close to 1.

But the problem is that full-scale investigation of the nature of superconductivity necessitates use of single crystals. At the same time, according to the Fe-Se system phase diagram [12], the FeSe tetragonal phase does not exist at temperatures above 457 ºC. The mass transfer process required for single crystal growing is hard to be organized at such a low temperature. In this work, we tried to grow FeSe single crystals by the gas-transport reactions method in order to shed light on the relation between the state of the material and its superconducting parameters, the critical $T_c$ temperature, in the first place.

Single crystals $FeSe_x$ were successfully grown earlier using a NaCl/KCl flux [13] and with application of the vapor self-transport method [14]. These crystals have a superconducting onset transition temperature $T_{co}$ ~ 11-12 K, which is notably higher than the polycrystalline value of $T_{co}$ ~ 8-9 K. In this work we present an observation of an even higher value of $T_{co}$ ~ 20-24 K.

Single crystals of $FeSe_x$ were grown using polycrystalline $FeSe_x$ by gas-transport reaction with $I_2$ as a gas-carrier. Ceramic $FeSe_x$ samples were prepared by synthesis in an ampoule at a temperature below 450 ºC, since, according to the Fe-Se system phase diagram [12], the temperature of decay for FeSe phase with tetragonal structure of PbO type is 457 ºC. Such a low temperature of synthesis requires a long phase equilibrium time, therefore after annealing the material was homogenized with intermediate grinding and pressing. As it was found out, only after four cycles, one week long each, the material presented one single phase. Because of such a long and complex procedure of synthesis, the material composition deviated from nominal mainly due to selenium evaporation; the evaporated selenium may deposit on the ampoule internal surface. The composition of ceramic samples was controlled by energy-dispersive analysis using the JEOL-733 microscope. It should be noted that this method of analysis being applied even with etalons, always gives a systematic error related to overestimation of the heavy elements content in comparison with lightweight elements. The value of this error was taken into corrected using $TiSe_2$ single crystals. For this compound, there is no homogeneity range with selenium excess [15], so the composition of most stoichiometric $TiSe_2$ samples was attributed to nominal composition. As the atomic weight of Fe is higher than that of Ti, such a correction also yields a rough estimate only, but it decreases the error in determining composition to a value less than the error caused by the polycrystalline state of the material and, respectively, by presence of grain boundaries, surface defects etc.

Phase composition of $FeSe_x$ ceramic was controlled by powder X-ray diffraction using the Shimadzu XRD 7000 diffractometer (Cu $K_\alpha$-radiation, bent monochromator, reflection mode, $2\vartheta$ range 8÷90 degrees). Work was carried out at the Ural-M center of the Institute of Metallurgy, UB RAS. According to diffraction data, $FeSe_x$ with tetragonal structure of PbO-type was the prevailing phase in the samples, the content of impurity phases was not more than 5

% (Fig. 1). The concentration dependence of unit cell parameters and unit cell volume is shown in Fig. 2. No traces of equilibrium two-phase region related to formation of $FeSe_x$ with hexagonal structure of NiAs-type were found in the studied concentration range. The change of unit cell parameters over the entire studied concentration range also points to absence of the high-temperature FeSe phase.

Crystals were grown in quartz ampoules with a temperature gradient from 700 ºC at the hot end to 300 ºC at the cold end of the ampoule; the ampoule length was 200-250 mm. The growing time was from 2 to 4 weeks. The typical obtained crystals were ~ 1 mm in size, and had a different composition which depended on the position of the crystal in the ampoule, and respectively, on the temperature of growth. The general tendency was increase of Se content at decrease of the temperature of growth. The crystals grown at the hot end of the ampoule were had close to $FeSe_{0.9}$ in their content but were inhomogeneous and contained some amount of metallic iron and $FeSe_x$ phase with tetragonal structure, while the main phase of these crystals had a hexagonal structure. Possibly, this composition was a result of non-equilibrium processes accompanying the decay of high-temperature $FeSe_x$ phase at cooling.

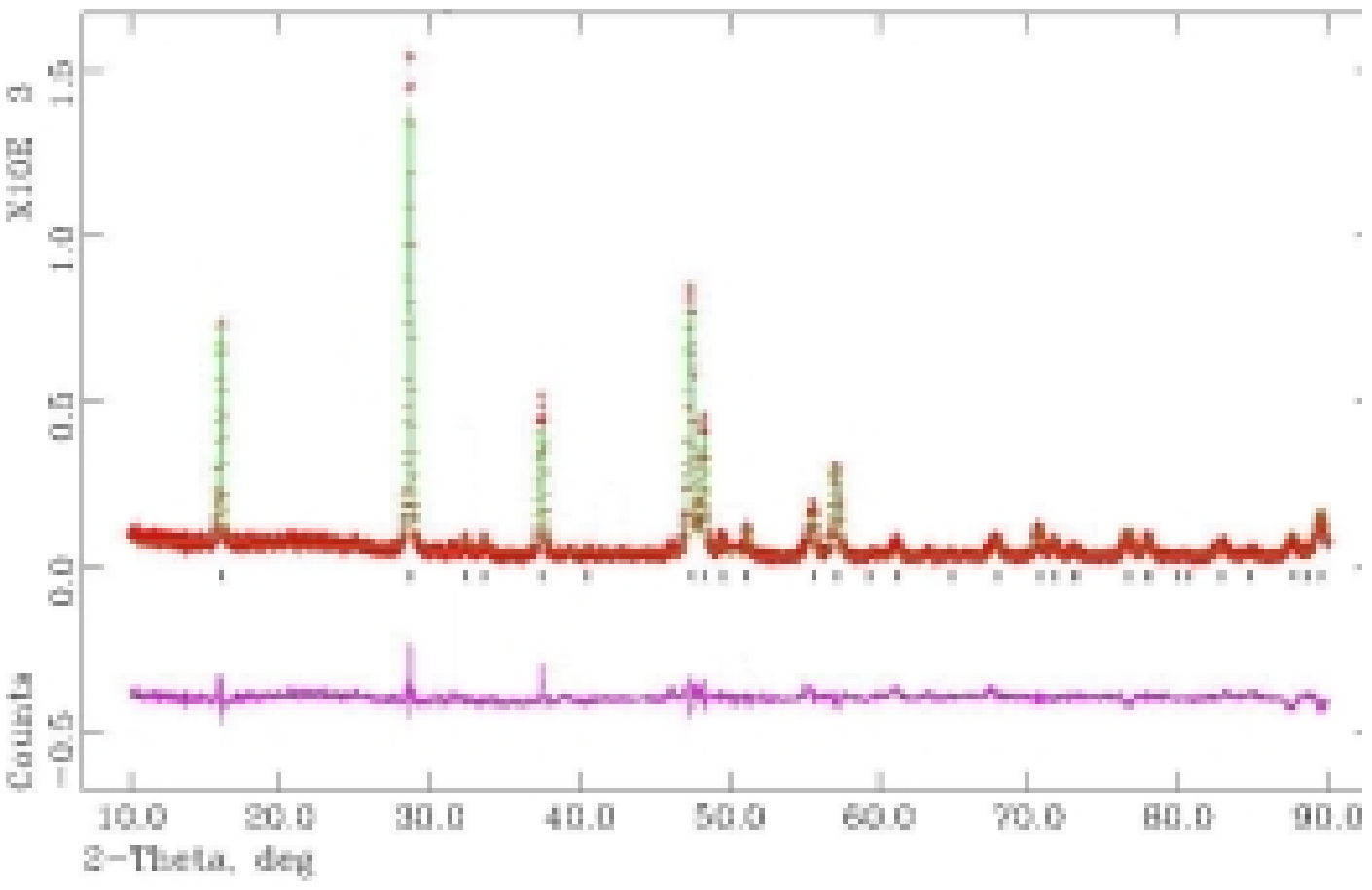

Fig.1 Typical diffraction pattern of a polycrystalline sample with nominal composition $FeSe_{0,94}$, consisting of tetragonal FeSe. Fitting was performed using the GSAS package, shown below is the difference curve.

The greater part of crystals was poorly fixed on the ampoule walls and formed a relatively homogeneous mixture at its opening. In this mixture, crystals with hexagonal-prismatic, tetragonal-prismatic, planar-square and hexagonal faceting could be recognized.

Resistivity $\rho$ was measured using the standard four-point method in magnetic field up to 13.6 T; magnetic measurements were made in the Quantum Design SQUID magnetometer.

Measurements showed the presence of a superconducting transition in crystals with all habit of crystal types (Fig. 3).

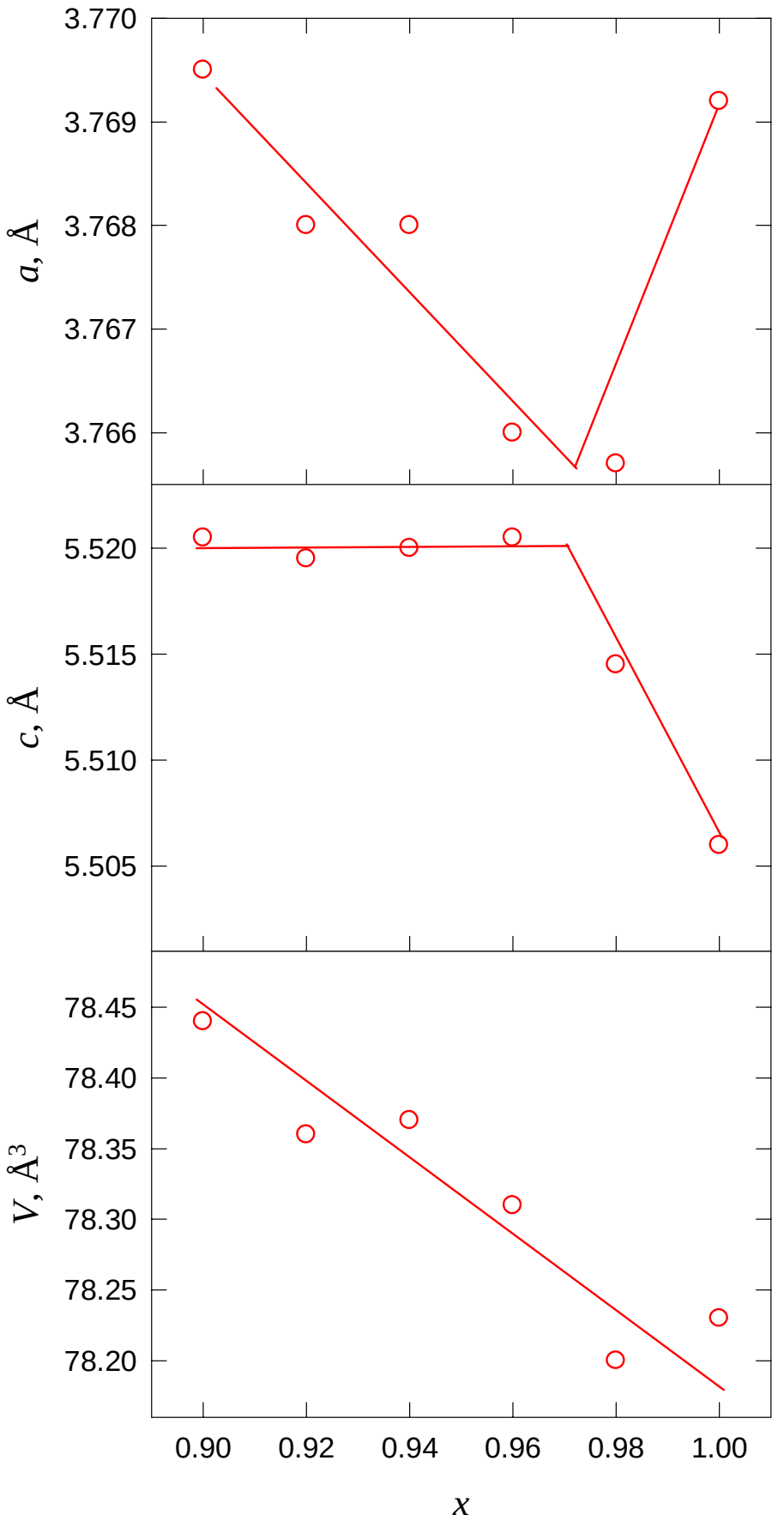

Fig. 2. Unit cell parameters *a*, *b* and unit cell volume *V* for polycrystalline samples of $FeSe_x$ as a function of *x*.

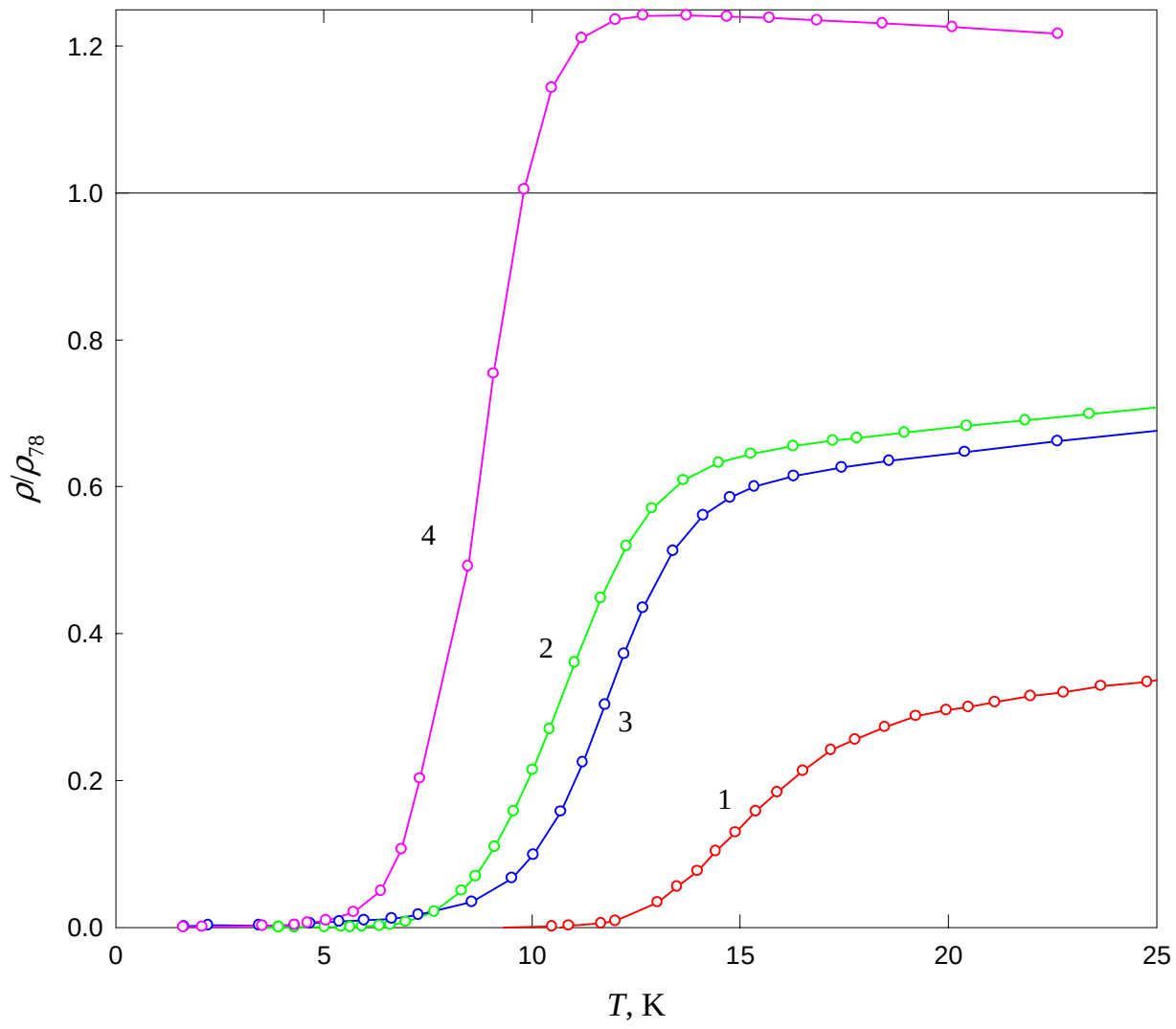

Fig. 3. Temperature dependences of reduced resistivity $\rho/\rho_{78}$ of $FeSe_x$ single crystals: 1 – tetragonal-prismatic; 2 – planar-square; 3 – planar hexagonal; 4 – hexagonal prismatic.

EDAX-analysis of crystals carried out in the QUANTA-200 scanning electron microscope showed the presence of heavy concentrated inhomogeneities. Apparently, such a situation may be due to nonequi-librium decay of the δ-FeSe high-temperature phase with formation of products of different composition.

Analysis of superconducting transition resistive curves allowed all crystals to be divided in two groups. The first group (see Fig. 3) demonstrates relatively narrow transitions, with the temperature $T_{cm}$ ~ 9-14 K was defined by the middle of transition and onset temperature $T_{co}$ ~12-18 K corresponded to the beginning of transition, which is generally in agreement with the available literature data [13, 14]. The second group (Fig. 4) shows wider transitions at $T_{cm}$ ~ 10-15.5 K and $T_{co}$ ~ 20-24 K. We found no literature data on the possibility of existence phase in Fe-Se system shows transition at $T_{co}$ of 20-24 K under normal pressure. As distinct from the first group of crystals, for which resistivity $\rho$ becomes zero at temperatures of 5-10 K, the value of $\rho$ for the second group remains finite at $T \geq 1.5$ K, which explicitly points to a low content of the superconducting phase.

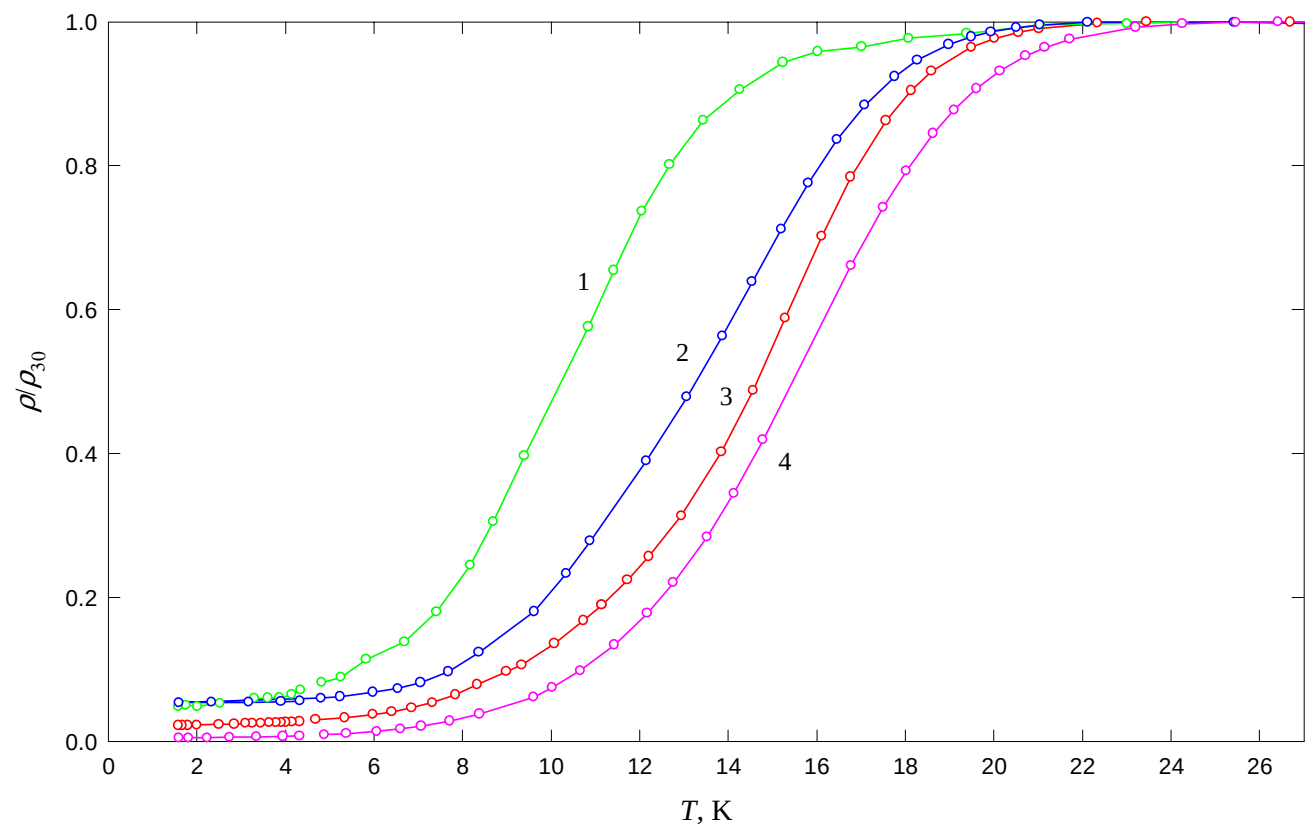

Fig. 4. Temperature dependences of reduced resistivity $\rho/\rho_{30}$ of a number of $FeSe_x$ samples with temperatures of beginning of superconducting transition $T_{co}$ ~ 20-24 K.

Resistive curves measurements in magnetic fields $H$ have demonstrated that, under growing $H$, resistive curves shift towards low temperatures without noticeable change in transition width, at an average rate of $-dH_{c2}/dT$ ~ 2 T/K (Fig. 5), so there is no reason to doubt that $T_{co}$ ~ 20-24 K is indeed related to the beginning of the onset of superconductivity.

Measurements of magnetic susceptibility also point to low superconducting phase content in samples of the second group (Fig. 6). Only at $T \leq 10$ K a noticeable drop of $\chi(T)$ is observed, which reaches only ~ 1.5 % of the screened phase (ZFC curve) at $T$ ~ 2 K. It may also be seen in the region of $10 < T < 20$ K that a probably small difference exists between the FC and ZFC curves which corresponds to concentration of the phase screened by superconducting currents of at least 0.1 %.

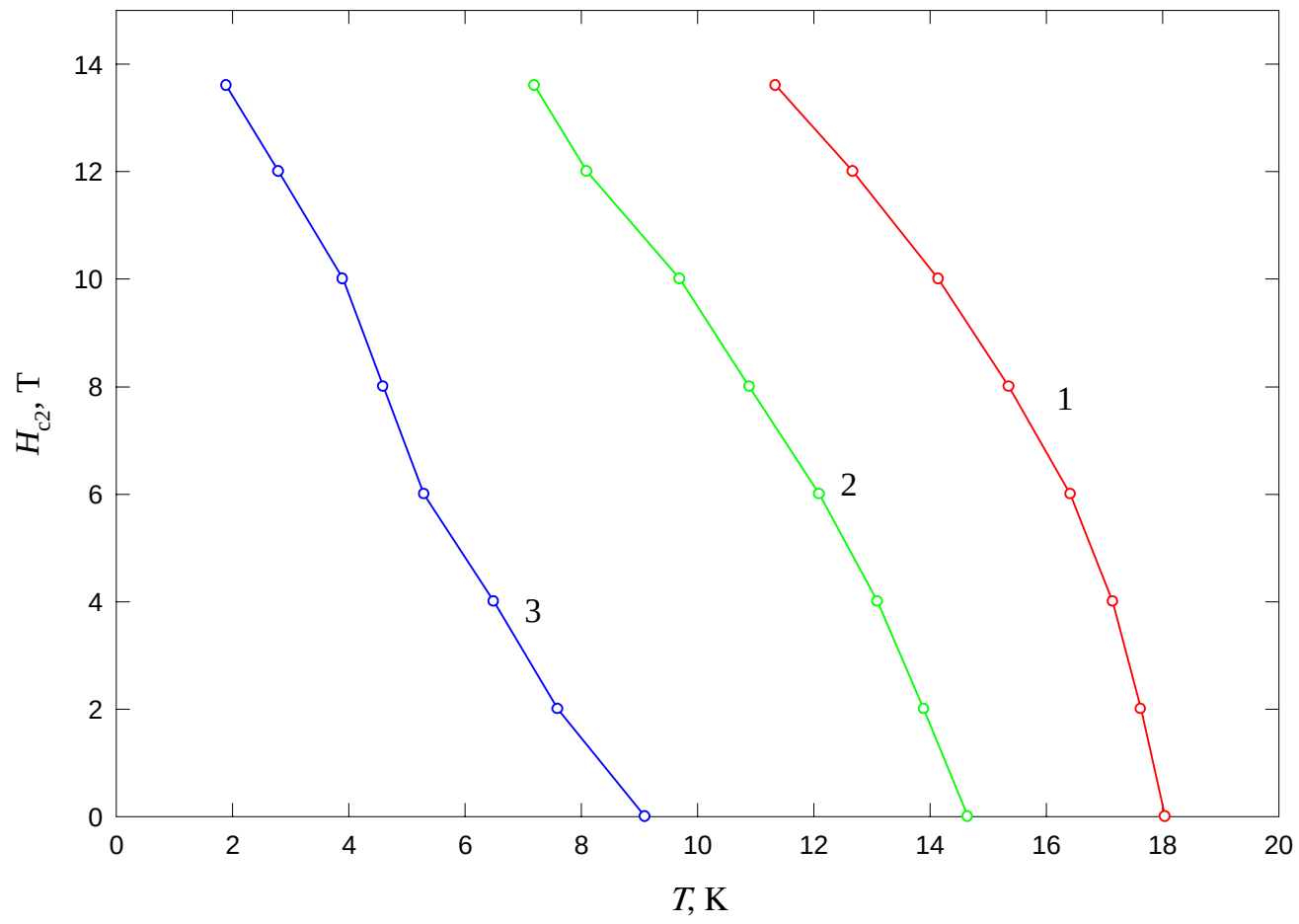

Fig. 5. Temperature dependences of resistive upper critical field $H_{c2}(T)$ defined at 90 % (1), 50 % (2) and 10 % (3) of normal state resistivity for sample 3 shown in Fig. 2.

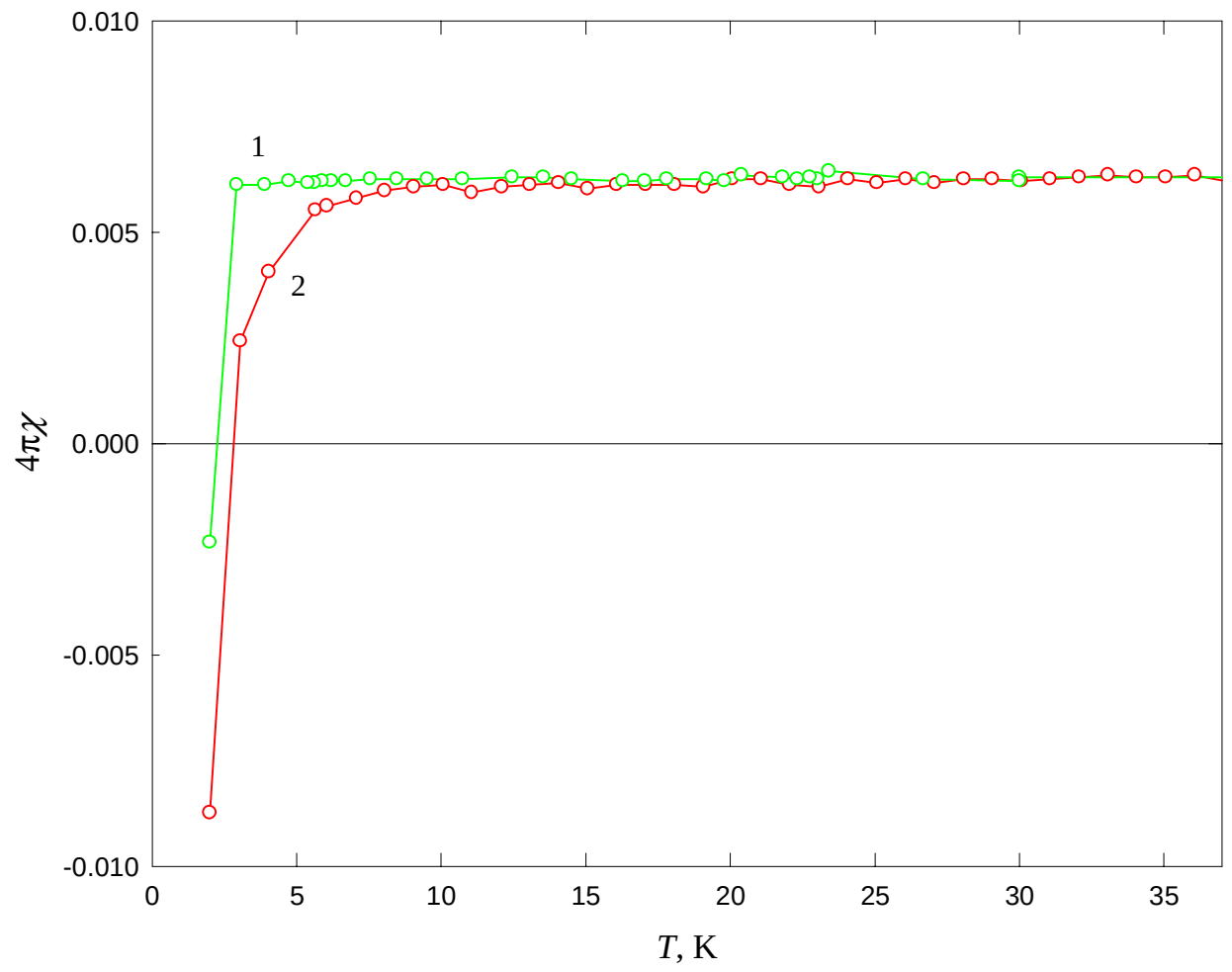

Fig. 6. Temperature dependences of magnetic susceptibility $\chi$ in magnetic field $H$ = 0.01 T in FC (1) and ZFC (2) modes for sample 3 shown in Fig. 2.

Such a discrepancy between magnetic and resistive data (less than 0.1 % of screened phase and almost complete resistive transition at $T > 10$ K) is probably due to the fact that the superconducting phase spatial configuration has the form of threads (or a network) of very small average volume, while the principal bulk of the samples presents a non-superconducting phase.

To make clear the possible stoichiometry of the phases with different $T_c$, we carried out annealing of non-superconducting crystals in gaseous contact with metallic copper. This was done with the help of a glass two-chamber non-sealed-off reactor. Crystals of FeSe were held at 400 ºC, the temperature of metallic copper was 300 ºC. As is known, copper activity at this temperature is higher than that of iron in the $Cu_2Se$ – FeSe system [16]. So, annealing in gaseous contact was expected to result in depletion of FeSe crystals in selenium. Really, after a week's annealing under the above conditions, metallic iron precipitates were registered on FeSe crystals surface. This allowed us to suggest that at least the surface part of FeSe crystals had acquired a composition corresponding to equilibrium with metallic iron. X-ray diffraction analysis of crystals (Fig. 7) ground to powder showed disappearance of hexagonal phase lines observed in the as-grown FeSe crystals. Apparently, this is connected with the decrease in selenium content.

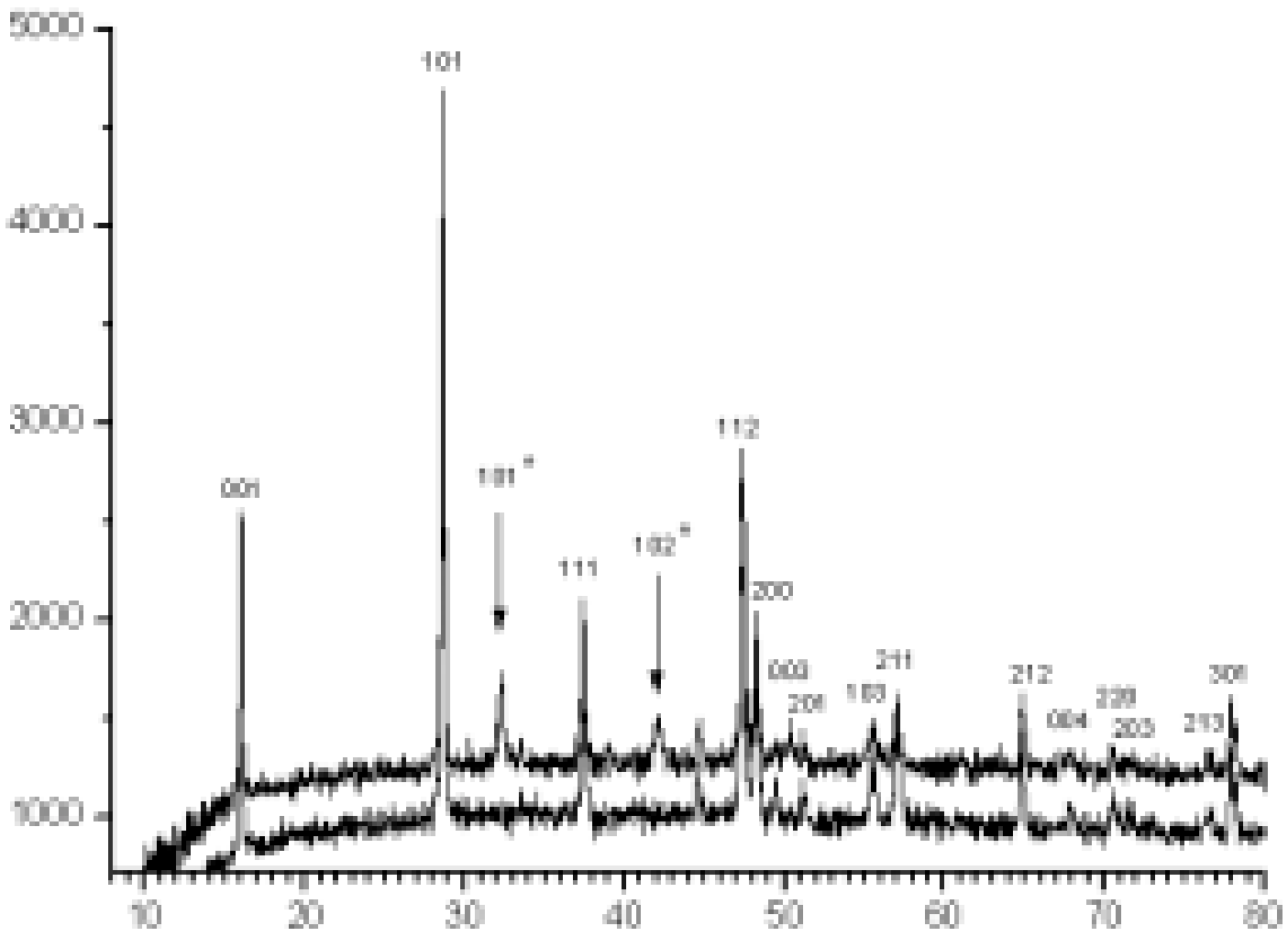

Fig. 7. Diffraction patterns of FeSe single crystals: as-grown (upper) and after a week's annealing in gaseous contact with metallic copper. Arrows point to FeSe hexagonal lines disappearing after annealing.

The temperature dependence of conductivity of all annealed single crystals demonstrated superconducting transition at $T_{cm}$ ~ 11 K and $T_{co}$ ~ 14 K. This is in good agreement with the first group crystals and the literature data [13, 14].

It is still unclear to which phase the transition at $T_{co}$ ~ 20 – 24 K may be related, since it is extremely difficult to identify a phase of such a small content. Note that ZFC curves give the upper concentration limit of the superconducting phase only, so the true content of the phase with high $T_{co}$ may be considerably below 0.1 %. The concentration dependence of $FeSe_x$ lattice parameters (see Fig. 2) points to the presence of two regions differing in that, in the region of $x = 0.9 – 0.98$, with growth of selenium deficit and at parameter $c_0$ staying constant, an increase in parameter $a_0$ is observed, while in the region of 0.98 – 1.0, parameter $c_0$ growth takes place, with simultaneous compression of parameter $a_0$ with decrease of $x$. Such behaviour is in agreement with the available literature data [7] and may point to difference in the nature of defects structure. For example, in the region of 1.0 –0.98, we may expect filling of interlayer spaces between FeSe layers, with simultaneous growth of defect structure in the layers proper. While in the region of 0.98 – 0.90, filling of defects within a layer with maximum possible filling of interlayer spaces takes place. It may be so that these two different types of defect structure may be responsible for formation of material with different $T_c$. In fact, as it may be seen in Fig. 2, the region of 1.0 – 0.98 corresponds to minimum cell volume. Clearly, the external pressure compressing the cell volume must stabilize exactly this state. At the same time, it is known that external pressure raises $T_c$ to a value of 27 – 37 K, which is close to the value of $T_{co}$ ~ 20 – 24 K we observed. Hence it is possible that the change in material composition having the same consequences as the application of external pressure will bring about similar changes in $T_c$. It is worth notice here that change in the material composition influences $T_c$ not through the change in the charge carriers concentration, but through deformation of the lattice, thus bringing it to the state of non-equilibrium (here, in the process of single crystals growth), and therefore must be related to the complex structure of the Fermi surface in this material.

As a conclusion, it is shown that grown crystals may be divided on two groups with “low” and “high” temperature of superconducting transition $T_c$. The crystals with low $T_c$ value are in equilibrium with metal iron and have maximal unit cell volume. The crystals with high $T_c$ value are enriched by selenium and have smaller unit cell volume. So, decrease of unit cell volume is accompanied by increase of $T_c$ value. As en external pressure leads to the same result [17], so we assume that unit cell volume rather than concentration of charge carriers determines the temperature of superconducting state transition for $FeSe_x$.

This work was carried with partial support of the Program of Basic Research of the Presidium of RAS “Condensed Matter Quantum Physics” (Project No. 4 UB RAS) and RFBR Projects No. 07-02-00020-a and 09-03-00053a. Authors are grateful for A. E. Kurmaev for helpful discussion.